# PacMap: Transferring PacMan to the Physical Realm

Thomas Chatzidimitris, Damianos Gavalas and Vlasios Kasapakis

Department of Cultural Technology and Communication University of the Aegean
Mytilene, Greece
`{tchatz, dgavalas, v.kasapakis}@aegean.gr`

**Abstract.** This paper discusses the implementation of the pervasive game *PacMap*. Openness and portability have been the main design objectives for PacMap. We elaborate on programming techniques which may be applicable to a broad range of location-based games that involve the movement of virtual characters over map interfaces. In particular, we present techniques to execute shortest path algorithms on spatial environments bypassing the restrictions imposed by commercial mapping services. Last, we present ways to improve the movement and enhance the intelligence of virtual characters taking into consideration the actions and position of players in location-based games.

## 1    Introduction

Pervasive gaming is an emerging gaming genre that transfers gameplay from the virtual world to the real environment, leading to the spatial, temporal and social expansion of the magic circle [1]. The key element in these games is the awareness and incorporation of user context: depending on the location, environmental or social context the game's scenario and the gameplay are adjusted accordingly.

When this genre of games appeared, the use of wearable devices (like sensors and GPS) was deemed necessary to capture user and environmental context, although the use of such equipment has been reported to affect the user's immersion during the gameplay. The advent of smartphones with their advanced processing, networking and sensory capabilities overturned the abovementioned restrictions of wearable equipment and provided pervasive games developers the means for implementing computationally intensive, context-aware applications commonly incorporating augmented reality.

This paper introduces PacMap, a pervasive variant of the classical game PacMan. PacMap has been largely inspired by Human PacMan [2], a milestone pervasive game project released in 2004. From the technology perspective, PacMap makes use of infrastructure and resources still unavailable at the time that Human PacMan was protoyped: it uses widely available equipment (like smartphones), 3G or WiFi networks, GPS and sensors. Furthermore, our prototype incorporates programming techniques and principles applicable to a wide range of location-based hunting/chase games. In particular, the implementation of PacMap aims at creating appealing and engaging game spaces, allowing anytime/anywhere gameplay, without any need for orchestration. The game stage is set around the actual location of the user and considers the actual surrounding road segments as game action 'corridors'. Moreover, we propose an implementation that utilizes the high level information granularity inherent in open map platforms and breaks the dependency on commercial map providers who set daily/monthly limits on the number of web service invocations. Finally, this paper suggests techniques for the smooth movement of virtual characters on map-based interfaces, which should adapt real-time on the actual player movement within the game space.

## 2      Related work

Location-based games claim a major share in pervasive games market. Many research prototypes [2] [3] [4] [5] [6] as well as some commercial games, like Ingress[1] and Zombies Run[2], use location-aware services to support their scenario, having the user's location as point of reference.

Human Pacman has been a milestone pervasive game (notably, one of the first to transfer the experience of an arcade game out to the physical world), which largely inspired the design of PacMap. Using a slightly modified game plot of the traditional Pacman, players are enrolled as pacmen, helpers and ghosts. The interaction, as well as the movement of players within the game space, requires the use of devices, like sensors and wireless LAN Cards, which are stored in a backpack. The players also carry head-mounted displays, whereon information about the plot of the game and augmented reality content are projected. The use of equipment, the need for orchestration (helpers) and the difficulty to set the game space at any location, seriously limits the portability and openness of the game.

The evolution of mobile computing (most notably, the emergence of mobile devices like smartphones, tablets, smart watches, etc), has radically changed the design and development of pervasive games. The incorporation of technologies, like GPS, sensors (accelerometer, gyroscope, proximity, compass, barometer, gesture, heart rate, etc) and built-in cameras provided game designers and developers the looked-for machinery to build location-based games with complex and appealing scenarios, and limited the requirement for specialized supplementary equipment. Notably, latest research prototypes commonly consume mapping services [5], like Google Maps, as well as relevant web services (e.g. directions for walking and transit transfers, points of interest, elevation, traffic and geocoding), which are provided by the service providers via specialized Application Programming Interfaces (APIs).

The use of several among the above mentioned services (e.g. the directions service) is subject to commercial usage. In practice, this limits the number of monthly invocations under a certain development license. This restriction raises a major challenge in the design of location-based games which involve heavy use of mapping services (e.g. chase/hunt games), especially those enrolling intelligent virtual characters and require execution of path finding algorithms for virtual hunters.

---

[1] https://play.google.com/store/apps/details?id=com.nianticproject.ingress
[2] https://play.google.com/store/apps/details?id=com.sixtostart.zombiesrun

## 3  Game scenario

The game scenario of PacMap adheres as much as possible to that of the classic arcade game Pacman. It is a location-based game, which requires Internet connection and enabled GPS receiver in the device. The game space is determined at startup, considering the actual road segments around the user's position as possible walking corridors for the pacman player. The user is supposed to collect all the cookies positioned across the streets. Unlike the pacman which us acted by a human player, the enemies (i.e. the ghosts) are virtual characters handled by the game engine. Similarly to the original arcade game, the ghosts are supposed to catch pacman, each following a different mobility pattern moving on the map, around the user's area. The purple, orange and blue ghosts execute random movements in the game space. The red ghost follows the user as the latter moves within the gamespace. Figure 1 illustrates a snapshot of the PacMap's gameplay.

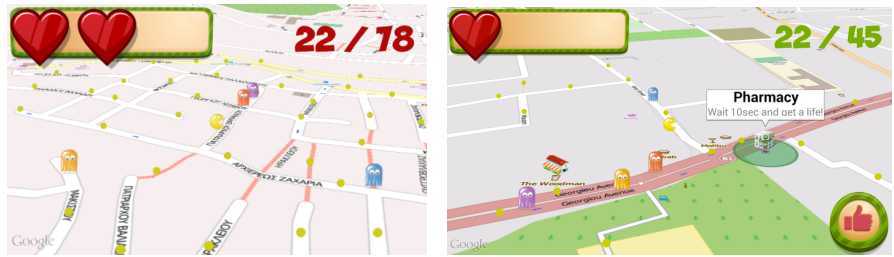

Figure 1. PacMap gameplay screenshots.

## 4  Game engine architecture

PacMap's system's architecture adopts a typical client-server model. The server side part undertakes the fabrication of the game space, whereas the client side visualizes the game space and enables the interaction among the player and the game engine.

As illustrated in Figure 2, the client sends out his location information to the game server in order to create the appropriate game space. The latter is confined by a circle around the user's location, with a radius of 200 meters. The game server uses the geolocation information to contact a map server and retrieve the nodes and POIs lying within this imaginary circle (the communication is handled by the OpenStreetMap API[3]).

---

[3] OpenStreetMap is an open-source mapping service, providing developers with useful crowdsourced topographical information. Data contributors may register geospatial elements

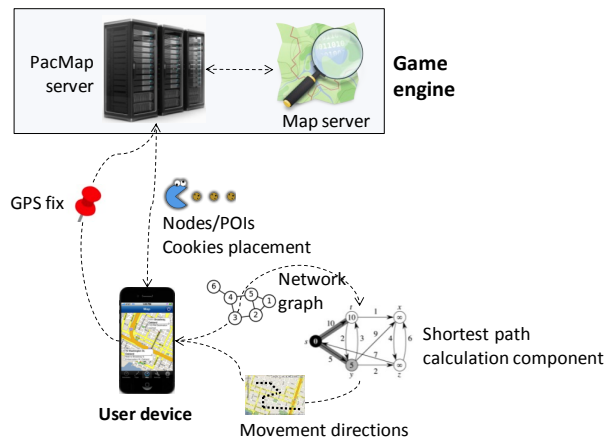

Figure 2.  PacMap system architecture.

The arrangement of game components within the game space (e.g. the placement of cookies) is carried out through utilizing the area nodes information. To ensure even distribution of cookies, the game engine firstly measures the distance between two nodes applying the Vincenty's formulae [7]. The latter is based upon two iterative methods, which are used in the field of Geodesy to measure the distance among two points on a spherical surface. Subsequently, the road sections are segmented, so that the cookies can be placed in equal distances.

In addition to nodes, POI information is also utilized in PacMap scenario, as the user can earn life credits, by reaching some of these POIs (like pharmacies and hospitals), or get "trapped" (e.g. in bar/nightlife areas) wherein the map visibility is reduced on the device's screen.

## 5     PacMap implementation

The arcade game Pacman involves ghosts which chase the pacman, with their movement promptly adapting to that of pacman. In order to transfer such functionality to a map-based interface each ghost needs to receive a series of road segments to be traversed. Provided a start and an end location (e.g. the current location of the ghost and the player, respectively) a reasonable action for the ghost following the user is to invoke a 'direction' web service typically offered by commercial map data providers (e.g. the Directions service of the Google Maps API) and then faithfully follow the shortest path walking directions recommended by the service. However, if the user location changes too often, direction service invocations (passing the updated ghost/player location parameters) will increase accordingly and soon exceed the invocations limit set by commercial providers.

---

such as nodes and POIs along the street network. Among others, the OpenStreetMap web services allow exporting the vertices of rectilinear parts, comprising a road network.

Enemies (i.e. ghosts) movement patterns fall into two types. Orange, blue and purple ghosts repeatedly execute a random movement around the map. To implement these movements, we derive two random pairs of coordinates over the arc of the imaginary circle centered at the user's current position. These two pairs of coordinates (representing the start/end nodes of each successive ghost movement) are submitted to the Directions API service of the map server, thus generating the actual path to be followed by each ghost. From the games research viewpoint, the movement of the red ghost is more challenging as it presumably applicable to many alike map/location-based chase games, since (according to the PacMap scenario) it is supposed to follow the user as s/he moves within the game space.

Ghosts movement respects the game space topography, namely the nodes exported from the game server during the game space generation phase (see Figure 3a). Considering a graph transformation of the game space (connecting adjacent nodes which are connected through a road segment on the actual setting and calculating the distances among them), it is then straightforward to execute a shortest path algorithm to compute the path to be followed by the user (see Figure 3b). PacMap's directions service implements the Dijkstra's algorithm[4], wherein edge costs equal the physical distance among their connected end nodes. For example, the red ghost considers the node nearest to the ghost's current location as start node and the node the user currently heads to as end node (e.g. node B in Figure 3a).

Shortest paths are derived whenever the user reaches a new edge or turns to another direction. To ensure prompt adaptation of ghost's movement to the player's movement, the device determines the nodes among which the user is currently located whenever his location (i.e. GPS fix) is updated.

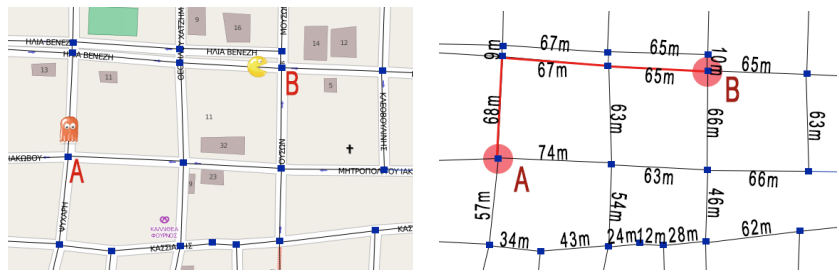

Figure 3. (a) An example game space illustrating the extracted topology nodes; (b) shortest path derived by a directions service executing Dijkstra's algorithm upon a graph transformation of the game space.

The sole goal of the ghost is to "catch" pacman, i.e to reach its currently assigned end node, having gone through the edge currently traversed by the user. In case that the ghost arrives at its end node without passing via the player's edge, the algorithm is re-executed, with the other end point of that edge set as the new end point for the requested directions.

It is noted that the shortest path algorithm is executed on the client side to eliminate the effect of network latency inherent in client-server interactions. On an average game space considering 420 nodes, the algorithm takes 95 msec to yield the shortest path when executed on a Samsung Galaxy S4 device (processor: ARMv7 – 1.8 GHz x 4 cores / ram:1.8 GB),

---

[4] Dijkstra's algorithm is a graph search algorithm that solves the single-source shortest path problem for a graph with non-negative edge path costs, producing a shortest path tree.

## 6    Conclusion

We introduced the prototype pervasive game PacMap, one of the few attempts to migrate a classical arcade game onto the physical realm. We have proposed programming techniques largely applicable to nearly any map-based chase game scenario, the main objective being to ensure openness and portability. We have also discussed implementation techniques for path-finding on real urban settings which bypass the restrictions imposed by commercial Direction APIs. The use of those techniques enables programmers and designers to develop location/map-based games, with flexible scenarios that involve intelligent virtual characters dynamically adapting on players movement behavior during the game.